\documentclass[aip,jcp,reprint]{revtex4-1}
\usepackage[applemac]{inputenc}
\usepackage{amssymb}
\usepackage{amsmath}
\usepackage{times}
\usepackage{graphicx}
\usepackage{MnSymbol}
\usepackage{subfigure}
\usepackage{natbib}
\usepackage{mciteplus}
\usepackage{dcolumn}
\usepackage{bm}
\usepackage[export]{adjustbox}
\usepackage{mathdots}
\usepackage{multirow}
\usepackage{epstopdf}
\usepackage{siunitx}
\usepackage{dcolumn}
\newcolumntype{d}{D{.}{.}{6} }

\begin{document}

\title{Elastic constants of Ice I$_h$ as described by semi-empirical water models}

\author{Pedro Augusto Franco Pinheiro Moreira}
\email{pmoreira@df.ufscar.br}
\affiliation{Departamento de F\'isica, UFSCar, Rodovia Washington Luiz, km 235, CP 676, 13565-905, S\~ao Carlos, S\~ao Paulo, Brazil}
\author{Roberto Gomes de Aguiar Veiga}
\email{roberto.veiga@ufabc.edu.br}
\affiliation{Centro de Engenharia, Modelagem e Ci\^encias Sociais Aplicadas, Universidade Federal do ABC, 09210-580, Santo Andr\'e, S\~ao Paulo, Brazil}
\author{Maurice de Koning}
\email{dekoning@ifi.unicamp.br}
\affiliation{Instituto de F{\'i}sica Gleb Wataghin, Universidade Estadual de Campinas, UNICAMP, 13083-859 Campinas, S{\~a}o Paulo, Brazil}
\affiliation{Center for Computational Engineering \& Sciences, Universidade Estadual de Campinas, UNICAMP, 13083-861, Campinas, S{\~a}o Paulo, Brazil}

\begin{abstract}

Using molecular dynamics simulations we compute the elastic constants of ice I$_h$ for a set of 8 frequently used semi-empirical potentials for water, namely the rigid-molecule SPC/E, TIP4P, TIP4P2005, TIP4P/Ice and TIP5P models, the flexible-molecule qTIP4P/Fw and SPC/Fw models and the coarse-grained atomic mW potential. In quantitative terms, the mW description gives values for the individual stiffness constants that are closest to experiment, whereas the explicit-proton models display substantial discrepancies. On the other hand, in contrast to all explicit-proton potentials, the mW model is unable to reproduce central qualitative trends such as the anisotropy in Young's modulus and the shear modulus. This suggests that the elastic behavior of ice I$_h$ is closely related to its molecular nature, which has been coarse-grained out in the mW model. These observations are consistent with other recent manifestations concerning the limitations of the mW model in the description of mechanical properties of ice I$_h$.

\end{abstract}

\date{\today}

\maketitle

\section{Introduction}
\label{sec1}

Due to its singular importance to life on Earth, the development of molecular-level descriptions for the condensed phases of water has been a very active area of investigation over the past 50 years. With the spectacular growth of available computational resources during this period, it has been characterized by a substantial increase of complexity in the search for models capable of describing a wide variety of thermodynamic and kinetic properties of both liquid and solid forms of water. In addition to the development of better exchange-correlation functionals for first-principles density-functional-theory (DFT) calculations,~\cite{Sun2016,Peng2016,Chen2017} substantial attention has also been given to the construction of semi-empirical models for water,~\cite{Guillot2002,Vega2011,Izadi2014,Cisneros2016} which are computationally much less demanding and allow the study of larger systems over longer time intervals.

The development of such analytical potential-energy functions (PEFs) for the description of water-water interactions has been driven predominantly by the goal of improving the corresponding structural, dielectric, thermodynamic and kinetic properties of the condensed phases of water.\cite{Mishima1998,Abascal2005,Vega2005a,GarciaFernandez2006,Vega2008,Handel2008,Liu2009,Molinero2009,MacDowell2010,Kastelowitz2010,Aragones2011,Limmer2011,Moore2011,Moore2011a,Johnston2012,Liu2012,Shepherd2012,Limmer2013,Sanz2013,Nguyen2015,Zaragoza2015,Cisneros2016,Espinosa2016,Espinosa2016a,Espinosa2016b,Lupi2017} The mechanical response characteristics, on the other hand, have remained mostly under the radar, with very few exceptions.~\cite{GelmanConstantin2015,Wu2015b,Min2018,Santos-Florez2018} This is somewhat surprising given that fundamental mechanical response functions such as the elastic constants directly mirror the underlying interactions between the constituent atoms or molecules that provide the cohesion in condensed phases. Accordingly, a proper description of elastic constants is a key element in the development of transferable PEFs and this is indeed a common goal pursued in the construction of semi-empirical models in general.~\cite{Martinez2013} For water PEFs, however, this has not been the case and their elastic constants are essentially unknown. Not only is their knowledge essential when choosing a model to study the mechanical behavior of solid phases of water,~\cite{Santos-Florez2018} they are also relevant, for instance, in the context of the melting behavior of a given PEF, which is known to correlate with elastic properties.~\cite{Burakovsky2003,Foata-Prestavoine2007}

The purpose of this paper is to compute the elastic constants for the most important crystalline phase of water on Earth, the proton-disordered hexagonal phase of water ice I$_h$,~\cite{Petrenko1999} for a set of 8 frequently used water PEFs, namely the rigid-molecule SPC/E,~\cite{Berendsen1987} TIP4P,~\cite{Jorgensen1985} TIP4P2005,~\cite{Abascal2005a} TIP4P/Ice~\cite{Abascal2005} and TIP5P models,~\cite{Mahoney2000} the flexible-molecule qTIP4P/Fw~\cite{Habershon2009} and SPC/Fw~\cite{Wu2006} PEFs and the coarse-grained atomic mW potential.~\cite{Molinero2009} For this purpose we carry out molecular dynamics (MD) simulations to determine the stress response to cell deformations under specified conditions of temperature and pressure. Using this approach we compute the elastic stiffness constants under a variety of thermodynamic conditions and compare the results to available experimental data. The remainder of the paper has been organized as follows. Sec.~\ref{sec2} describes the theoretical background and the adopted computational methodology, followed by a discussion of the results in Sec.~\ref{sec3}. Finally, we conclude with a summary in Sec.~\ref{sec4}.

\section{Computational approach}
\label{sec2}
The elastic stiffness of a substance is described by a rank-four tensor with components $C_{ijkl}$ that establishes the relation between the state of deformation, specified by the rank-two strain tensor $\varepsilon_{ij}$ and the corresponding applied stress tensor $\sigma_{ij}$, where the suffixes $i,j,k$ usually represent the Cartesian directions $x$, $y$ and $z$. In the linear regime this relation is defined by~\cite{Nye1985}
\begin{equation}
\sigma_{ij} = \sum_{\substack{kl= \\x,y,z}} C_{ijkl} \varepsilon_{kl}.
\label{eq:stiffness}
\end{equation}
Similarly, the definition of the elastic compliance tensor with components $S_{ijkl}$ is given by
\begin{equation}
\varepsilon_{ij} = \sum_{\substack{kl= \\ x,y,z}} S_{ijkl} \sigma_{kl}.
\label{eq:compliance}
\end{equation}

Due to the symmetries, these rank-4 tensors can be represented in terms of $6\times 6$ matrices by using Voigt notation,~\cite{Nye1985} in which the first and second pairs of suffixes, i.e., $ij$ and $kl$, are represented by a single index running from 1 to 6 according to the relations $(11) \to 1$, $(22)\to 2$, $(33) \to 3$, $(23),(32) \to 4$, $(13),(31) \to 5$ and $(12),(21) \to 6$. In the same fashion the rank-2 stress and strain tensors are represented by column matrices with 6 lines. Accordingly, Eq.~(\ref{eq:stiffness}) can be represented as the matrix-vector multiplication
\begin{equation}
\sigma_{k}= \sum_{j=1}^6 C_{kj}\,\varepsilon_j.
\label{eq:stiffness_voigt}
\end{equation}
An analogous matrix representation $S_{kj}$ exists for the elastic compliance tensor defined in Eq.(\ref{eq:compliance}). Moreover, its elements can be determined by inverting the matrix formed by the stiffness constants $C_{kj}$.

There are several ways in which elastic stiffness constants can be computed from atomistic simulations.~\cite{Ray1988} Here, we adopt the direct approach~\cite{Ray1988,Jong2015} in which the cell is subjected to a deformation in which only one of the six independent strain components, say $\varepsilon_l$, is nonzero. In this manner, there is only one term in the summation of Eq.~(\ref{eq:stiffness_voigt}) for each stress component, i.e.,
\begin{equation}
\sigma_k=C_{kl}\varepsilon_l, \ \ \ \ k = 1,\cdots,6.
\end{equation} 

\begin{figure}[ht!]
\centering
  \includegraphics[width=8.6cm]{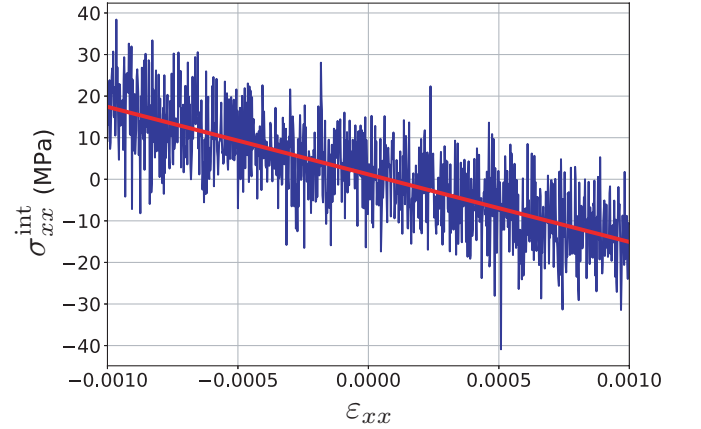}
  \caption{Typical stress-strain relation for deformation simulations of ice I$_h$. Blue line displays the internal $\sigma_{1}=\sigma_{xx}$ stress response associated with the tensile $\varepsilon_{xx}=\varepsilon_{1}$ deformation as described by the TIP4P/Ice model at a temperature of $T=235.5$~K and zero pressure. The $x$ direction corresponds to the $[0\bar110]$ crystallographic axis. Red line is linear fit to stress-strain data and its slope corresponds to elastic stiffness constant $C_{11}$. The statistical uncertainty in the slope gives the error bar in the elastic-constant estimate.}
  \label{Fig1}
\end{figure}

In practice, we explore this linear relationship by monitoring the internal stress during a dynamic deformation protocol in which the chosen strain component varies between $-\varepsilon_{\rm max}$ and $+\varepsilon_{\rm max}$ during a simulation. As long as the value of $\varepsilon_{\rm max}$ remains sufficiently small for linear elasticity to be valid, the elastic constants can be determined by fitting the corresponding stress-strain response to a straight line, with the slope,
\begin{equation}
C_{kl}=-\frac{d\sigma^{\rm int}_k}{d\varepsilon_l},
\end{equation}
where $\sigma^{\rm int}_{ij}$ is the internal stress and the minus sign reflects the fact that the relation between the internal and applied stress in a state of mechanical equilibrium is $\sigma_{ij}^{\rm int}=-\sigma_{ij}$. A typical example is shown in Fig.~\ref{Fig1}, which displays the internal $\sigma^{\rm int}_{xx}\equiv\sigma_{1}$ stress response associated with the tensile deformation $\varepsilon_{xx}\equiv\varepsilon_{1}$ for ice I$_h$ as described by the TIP4P/Ice model. The appreciable fluctuations are due to the elevated temperature of $T=235.5$~K. The $x$ direction corresponds to the $[0\bar110]$ crystallographic axis. The slope of the linear fit to the data, shown as the red line, gives the value of $C_{11}$. Similarly, fitting the other 5 stress components during the deformation simulation gives estimates for the remaining $C_{k1}$ with $k=2,\cdots,6$. Carrying out dynamic deformation simulations for all 6 independent strain components then gives estimates for the 36 elements of the stiffness matrix $C_{kj}$ in Eq.~(\ref{eq:stiffness_voigt}). 

The nature of the elastic stiffness constants and their corresponding compliances depends on the thermodynamic constraints during the deformation process. Two common cases are those in which the deformations are carried out isothermally and adiabatically, described by isothermal $(C^{T}_{kl})$ and adiabatic ($C^{S}_{kl}$) stiffness constants, respectively.~\cite{Wallace1998,Ray1986} While the isothermal elastic stiffnesses describe the stress-response to constant-temperature deformations, their adiabatic counterparts quantify the induced stress during isentropic deformations in which the system is decoupled from a heat bath. Given that for ice I$_h$ experimental results are mostly based on Brillouin spectroscopy~\cite{Gammon1980,Gammon1983,Gammon1983a,Gagnon1988} or sound-velocity measurements~\cite{Proctor1966}, both of which probe the adiabatic stiffness constants $C^{S}_{kl}$, our calculations focus on the latter.  

To compute the adiabatic stiffnesses at specified conditions of temperature $T$ and pressure $P$ we employ the following 3-step procedure. First, the undeformed system is equilibrated at the desired temperature and pressure within the $NPT$ ensemble, computing the corresponding equilibrium dimensions of the computational cell. Second, the cell is strained by an amount $-\varepsilon_{\rm max}$ according to one of the 6 deformation modes and it is equilibrated at temperature $T$ at fixed cell geometry. Third, the cell is decoupled from the heat reservoir, after which the deformation of the cell is ramped from the initial value $-\varepsilon_{\rm max}$ to $+\varepsilon_{\rm max}$ at a constant strain rate $\dot{\gamma}$. In this fashion, the magnitude of deformation along the simulation evolves according to
\begin{equation}
\varepsilon(t)=-\varepsilon_{\rm max}+\dot{\gamma}\,t,
\end{equation}
where $t$ is the physical time during the simulation. Accordingly, the duration of the deformation process is given by $t_{\rm tot}=2\,\varepsilon_{\rm max}/\dot{\gamma}$.

All calculations have been carried out using the Large-scale Atomic/Molecular Massively Parallel Simulator (LAMMPS)~\cite{Plimpton1995} MD package. The employed computational cells contain 1600 water molecules in a proton-disordered ice I$_h$ structure with zero total dipole moment.~\cite{Santos-Florez2018} For all explicit water models the long-range electrostatics is handled employing the particle-particle particle-mesh (PPPM) scheme.~\cite{Hockney2017} The bond lengths and angles for the rigid-molecule PFEs are constrained using the SHAKE algorithm.~\cite{Ryckaert1977} For the coarse-grained mW model, due to the absence of protons, neither SHAKE nor PPPM are required. 

Equilibration is achieved by running 1~ns-long simulations within an anisotropic zero-pressure isobaric-isothermal NPT ensemble in which all three sides of the orthogonal cell are allowed to vary independently. The corresponding equations of motion are based on a Parrinello-Rahman-type barostat~\cite{Parrinello1981,Martyna1994,Shinoda2004} and a Langevin thermostat.~\cite{Schneider1978} They are integrated using time steps of $\Delta t = 5.0$~fs, 2.0~fs and 0.5~fs, respectively, for the mW model, the rigid-molecule PFEs and the flexible-water descriptions. The damping time scales for the thermostat and barostat are chosen as $\tau_T=100 \Delta t$ and $\tau_P=1000 \Delta t$, respectively. During the deformation runs the Langevin thermostat is turned off. Furthermore, we use a maximum strain magnitude of $\varepsilon_{\rm max} = 0.001$ at a strain rate of $\dot{\gamma}=1\times10^6$~s$^{-1}$, which amounts to a total simulation time $t_{\rm tot}=2$~ns for each deformation process. This deformation rate is sufficiently slow for the results to be independent of its value.

For finite-temperature simulations such as those shown in Fig.~\ref{Fig1}, in which thermal fluctuations in the internal stress are appreciable, the stiffness constants are obtained by averaging the results over 20 independent straining simulations for each deformation mode.

\begin{table*}[th!]
\begin{center}
\caption{Zero-pressure and zero-temperature estimates of the five independent elastic stiffness components $C_{ij}$, Young's moduli $E_i$, the shear moduli $G_{ij}$ and bulk moduli \textit{B} (in GPa) for ice I$_h$ for the 8 considered water models. The subscripts $1$ and $2$ refer to any pair of perpendicular directions in the basal plane, while subscript $3$ is associated with the $c$-axis. First line contains extrapolated experimental estimates from Ref.~\onlinecite{Proctor1966}. Numbers in parentheses denote uncertainty in final digit.}
\label{tab:isostiff}
\begin{tabular}{| l | d | d | d | d | d | d | d | d | d | d |}
\hline
           & \multicolumn{1}{c|}{$C_{11}$} & \multicolumn{1}{c|}{$C_{12}$} & \multicolumn{1}{c|}{$C_{13}$} & \multicolumn{1}{c|}{$C_{33}$}  & \multicolumn{1}{c|}{$C_{44}$} & \multicolumn{1}{c|}{$E_{1}$,$E_{2}$} & \multicolumn{1}{c|}{$E_{3}$} & \multicolumn{1}{c|}{$G_{13}$,$G_{23}$} & \multicolumn{1}{c|}{$G_{12}$} & \multicolumn{1}{c|}{$B$}          \\
\hline
Expt. (Ref.~\onlinecite{Proctor1966})      & 17.10(1)  & 8.5(1)   & 7.13(4) & 18.21(1)  & 3.62(1) & 12.03(1) & 14.24(1) & 3.62(1)& 4.30(5)& 10.88(3) \\
mW   & 14.482(2) & 5.740(3)   & 5.067(2)  & 15.156(3) & 3.972(3) & 11.507(2) & 12.616(3) & 3.971(1) & 4.371(3) & 8.430(2) \\
SPC-E      & 22.170(6) &15.173(6) & 13.475(4) & 23.42(1)  & 3.030(3) & 10.597(6) & 13.69(1) & 3.030(3) & 3.498(4) & 16.89(3) \\
SPC-Fw     & 22.90(2) & 16.23(2)  & 14.41(1) & 24.30(3) & 2.861(8) & 10.24(2) & 13.69(3) & 2.861(8) & 3.34(1) & 17.8(1)\\
TIP4P      & 20.688(5) & 14.607(5) & 13.619(4) & 21.486(8)  & 2.763(3) & 9.095(6) & 10.977(8) & 2.763(3) & 3.041(4) & 16.28(4) \\
TIP4P-2005 & 22.271(5) & 15.802(5)  & 14.659(4) & 22.690(9)  & 2.858(3) & 9.669(5) & 11.402(9) & 2.858(3) & 3.235(4) & 17.49(4) \\
TIP4P-Ice  & 23.871(2) & 17.013(2)  & 15.779(2) & 24.105(3)  & 3.025(1) & 10.244(2) & 11.925(3) & 3.025(1) & 3.430(1) & 18.77(2)  \\
qTIP4P-F   & 22.03(2) & 15.98(2)  & 14.94(1) & 22.78(2)  & 2.695(7) & 9.11(2) & 11.03(2) & 2.695(7) & 3.03(1) & 17.6(1) \\
TIP5P      & 32.218(5) & 13.951(5)  & 10.873(4) & 35.455(9)  & 8.040(3) & 24.981(5) & 30.333(9) & 8.040(3) & 9.134(4) & 19.03(7) \\
\hline
\end{tabular}
\end{center}
\end{table*}

\section{Results and Discussion}
\label{sec3}
As a first step we estimate the fundamental elastic stiffness constants in the limit of zero temperature and pressure. To approach this limit, we equilibrate the systems at 0.01 $K$ and zero pressure before carrying out the adiabatic deformation protocol. The results are shown in Table~\ref{tab:isostiff}, which displays the 5 independent elastic stiffness constants, Young's moduli, the shear moduli and the bulk moduli for the 8 considered water models and compares them to extrapolated zero-temperature experimental estimates.~\cite{Proctor1966} 

A first conclusion is that all explicit-proton water models, both rigid and flexible, in strongly overestimate the stiffness constants related to tensile and compressive deformations. The overestimates are most pronounced for $C_{12}$, $C_{13}$ and the bulk modulus, for which these models give values that are up to twice as large as the experimental data. It is also noteworthy to observe that, even though the functional forms of the SPC and TIP4P-based rigid and flexible models are quite different, the variation in the elastic constants between them is of the order of only $\sim$10~\%. The TIP5P PEF is an exception, giving values for $C_{11}$ and $C_{33}$ that are $\sim$50~\% larger than those of the other explicit-proton models. In contrast to the tensile/compressive elastic-constant overestimates, the shear moduli are systematically lower than the experimental values, as can be seen in the columns with $C_{44}$ and $G_{12}$. Again, the TIP5P represents an exception, also strongly overestimating the shear moduli. Compared to the explicit-proton models, the coarse-grained mW model provides an overall better description of the individual stiffness constants, showing discrepancies with experiment that are at most $\sim 30$~\%.

The variation in the experimental values of Young's moduli $E_i$ and the shear moduli $G_{ij}$ for the different directions in Table~\ref{tab:isostiff} shows that ice I$_h$ is elastically anisotropic. Given that crystals with hexagonal symmetry are isotropic in the basal plane, this anisotropy can be described entirely by a single angular variable $\theta$ that measures the orientation with respect to the $c$-axis. In particular, the anisotropy in Young's and the shear moduli in ice I$_h$ are given by~\cite{Schulson2009}
\begin{equation}
E(\theta)=\left[S_{11} \sin^4\theta+S_{33}\cos^4\theta+(S_{44}+2S_{13})\sin^2\theta\cos^2\theta\right]^{-1},
\end{equation}
and
\begin{eqnarray}
\nonumber
G(\theta)&=&\left[S_{44} + (S_{11}- S_{12}-S_{44}/2)\sin^2\theta+\right. \\
&+&\left. 2(S_{11}+S_{33}-2S_{13}-S_{44})\sin^2\theta\cos^2\theta\right]^{-1},
\end{eqnarray} 
where the $S_{ij}$ are the elastic compliance constants.

\begin{figure}[t!]
\centering
  \includegraphics[width=8.6cm]{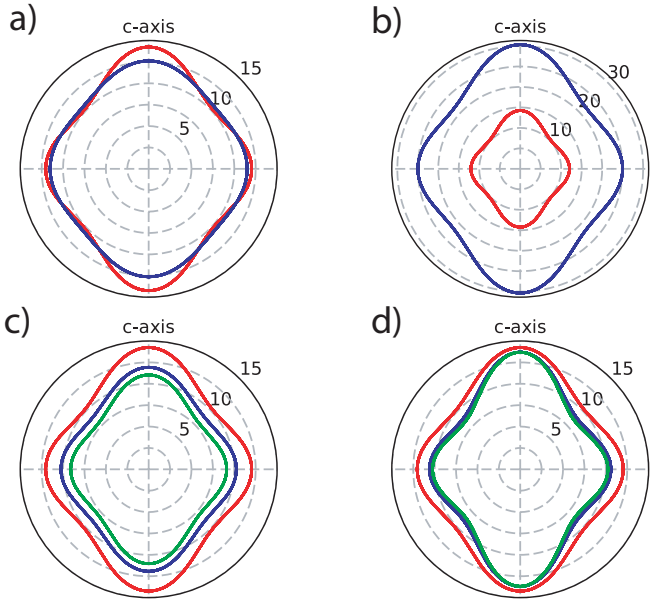}
  \caption{Polar plots comparing anisotropy in Young's modulus at zero pressure and temperature between experimental estimates from Ref.~\cite{Proctor1966} and semi-empirical models as obtained from results in Table~\ref{tab:isostiff}. Radial units are in GPa. Vertical direction corresponds to $c$-axis. Red lines represent experimental data. (a) mW model (blue line) (b) TIP5P model (blue line), c) TIP4P/Ice (blue line) and qTIP4P-F (green line), d) SPC/E (blue line) and SPC/Fw (green line). }
  \label{Fig2}
\end{figure}

\begin{figure}[t!]
\centering
  \includegraphics[width=8.6cm]{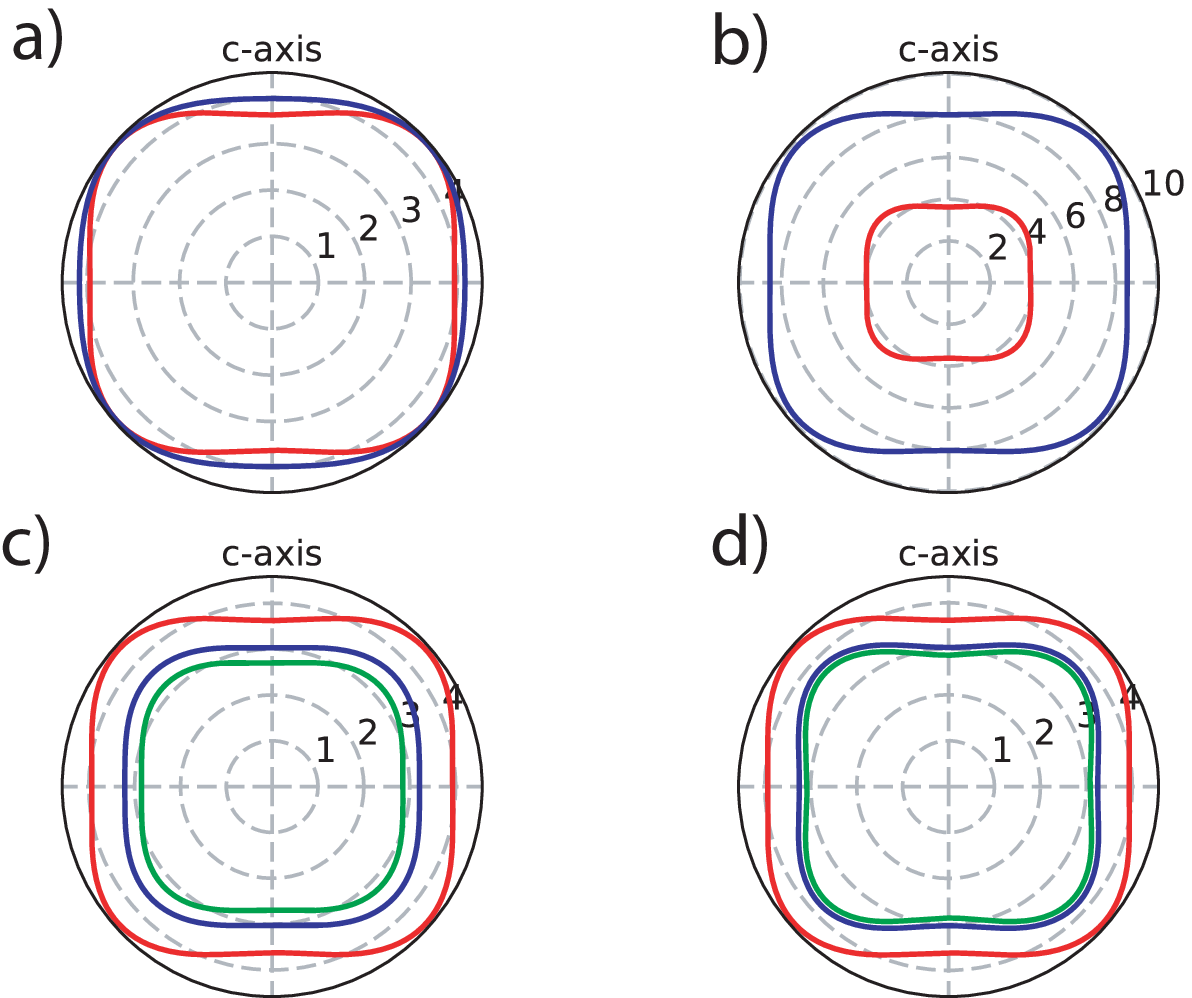}
  \caption{Polar plots comparing anisotropy in shear modulus at zero pressure and temperature between experimental estimates from Ref.~\onlinecite{Proctor1966} and semi-empirical models as obtained from results in Table~\ref{tab:isostiff}. Radial units are in GPa. Vertical direction corresponds to $c$-axis. Red lines represent experimental data. (a) mW model (blue line) (b) TIP5P model (blue line), c) TIP4P/Ice (blue line) and qTIP4P-F (green line), d) SPC/E (blue line) and SPC/Fw (green line).  }
  \label{Fig3}
\end{figure}

Figures~\ref{Fig2} and ~\ref{Fig3} show resulting profiles for a number of ice models at zero temperature and pressure and compares them to the corresponding experimental values obtained from Ref.~\cite{Proctor1966}. 
The experimental data for Young's modulus in Fig.~\ref{Fig2} show that its anisotropy, measured as the difference between maximum and minimum values relative to the minimum value, is of the order of $\sim20$~\%. As shown in Fig.~\ref{Fig2}a) the coarse-grained mW model substantially underestimates this anisotropy. Even though the absolute values of the Young's moduli are in quite good agreement, it does not reproduce the anisotropy profile and gives an anisotropy less than $\sim 10$~\%. All explicit-proton models, on the other hand, both rigid and flexible in nature, give anisotropies that closely resemble the experimental profile, as shown in Fig.~\ref{fig2} b), c) and d). Indeed, as displayed in Fig.~\ref{Fig2}b), even the TIP5P PEF, which quantitatively overestimates the absolute values of Young's modulus by a factor two, displays an anisotropy that is qualitatively very similar to experiment. 

The anisotropy results for the shear modulus, as portrayed in Fig.~\ref{Fig3}, show a comparable picture. The mW model gives shear moduli that are quantitatively the closest to experiment, but underrates the anisotropy. The explicit-proton models, on the other hand, describe an anisotropy that closely resembles the qualitative profile displayed by the experimental data, but predict quantitative shear-modulus values that deviate significantly from experiment. 

Next, we assess the temperature and pressure dependence of the adiabatic elastic stiffness constants for two of the most popular models in the description of ice I$_h$, namely the TIP4P/Ice and the mW models. To compare the results to available experimental data~\cite{Gagnon1988} we determine the temperature dependence at a hydrostatic pressure of 500 bar, as well as the pressure dependence of ice at 35.5 K below the melting temperature. 

\begin{figure}[t!]
\centering
  \includegraphics[width=8.6cm]{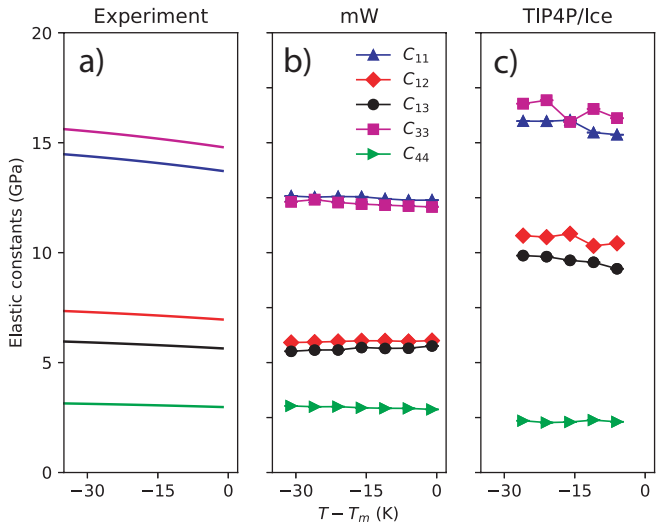}
  \caption{Temperature dependence of the adiabatic elastic stiffness constants at a hydrostatic pressure of 500 bar. For the mW and TIP4P/Ice models, the temperature values are chosen relative to the melting temperatures $T_m=276$~K and 271~K of the respective models. Lines in (a) represent the quadratic fits reported in Ref.~\onlinecite{Gagnon1988}. Data shown in (b) and (c) corresponds to the simulation results obtained for the mR and TIP4P/Ice models, respectively.}
  \label{Fig4}
\end{figure}

Fig.~\ref{Fig4} shows the temperature dependence of the adiabatic stiffness constants at 500 bar. The experimental data show a softening as the temperature increases. This thermal softening, which is a consequence of anharmonic effects,~\cite{Leibfried1961} is rather modest, however, being less than $\sim5$~\% over the interval between 30 and 5 K below the melting temperature $T_m$. Aside from quantitative differences in the stiffness-constant values both models generally reproduce the correct magnitude of the softening, with the exception of $C_{12}$ and $C_{13}$ for the mW model. These particular constants display a small but significant increase of $\sim1.5-3$~\% over the considered temperature interval. While uncommon, such thermal stiffening has been observed in a number of salts~\cite{Leibfried1961} and is possible only if the inter-particle interactions are of many-body (i.e., non-central) character of the inter-particle forces.~\cite{Leibfried1961} In this light the stiffening of $C_{12}$ and $C_{13}$ with increasing temperature is related to the three-body term in the mW model, but the disagreement with the experimentally established softening is a direct indication that it does not provide a correct description for the angular forces in ice I$_h$. 

\begin{figure}[t!]
\centering
  \includegraphics[width=8.6cm]{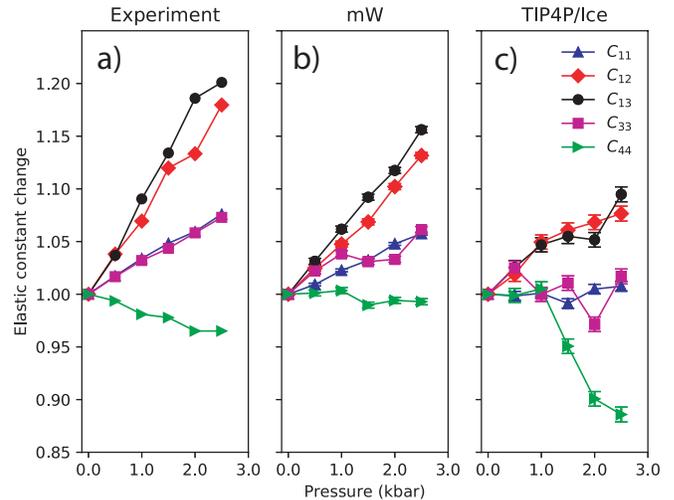}
  \caption{Pressure variation of the adiabatic elastic stiffness constants at 35.5 K below the $T_m$. Data is displayed relative to zero-pressure values. Panel (a) depicts the quadratic fits to experimental data reported in Ref.~\onlinecite{Gagnon1988}, whereas (b) and (c) depict simulation results for the mW and TIP4P/Ice models, respectively. For both PEFs, the melting-temperature reference was chosen to be the corresponding values  $T_m=276$~K and 271~K for the mW and TIP4P/Ice models, respectively. )  }
  \label{Fig5}
\end{figure}

Fig.~\ref{Fig5} depicts the change of the adiabatic elastic constants as a function of hydrostatic pressure at a temperature of 35.5 K below $T_m$. The dependence is plotted as the ratio of the finite-pressure value to that of the stiffness at zero pressure. The experimental data from Ref.~\onlinecite{Gagnon1988} show that the stiffness constants, with the exception of $C_{44}$, increase with pressure due to volume contraction. Indeed, $C_{12}$ and $C_{13}$ increase by as much 20~\% at a pressure of 2.5 kbar. The softening of shear modulus $C_{44}$, on the other hand, signals a reduction of the stability of the ice I$_h$ phase. This is consistent with the close proximity of the ice I$_h$-ice II and ice I$_h$-ice III coexistence lines under these conditions.~\cite{Chaplin} The mW model closely reproduces the experimental trends, giving the constants $C_{12}$ and $C_{13}$ as those that increase by the most, of the order of $\sim15$~\%, followed by $C_{11}$ and $C_{33}$ which stiffen by $\sim 6$~\%. There is, however, virtually no softening of $C_{44}$, which is possibly related to the absence of phase transitions in its phase diagram~\cite{Dhabal2016} in the considered range of pressure values. For the TIP4P/Ice model, on the other hand, $C_{44}$ softens by almost 10~\% upon increasing the pressure to 2.5 kbar. As for the experimental case, this is concordant with the vicinity of the I$_h$-ice II and ice I$_h$-ice III coexistence lines in TIP4P/Ice's phase diagram.~\cite{Abascal2005} The model also qualitatively matches the trends for the elastic constants $C_{11}$ and $C_{33}$ although the magnitude of the stiffening is roughly a factor 2 lower compared to experiment.

The results above show that, overall, the agreement between experimental values of the stiffness constants and model predictions is poor, as becomes clear from Table~\ref{tab:isostiff} and Figs.~\ref{Fig4} and \ref{Fig5}. The coarse-grained mW model most closely reproduces experimental values from a quantitative standpoint, both in the zero-temperature limit as well as for the considered finite-temperature and pressure behaviors, whereas the explicit-proton models display much larger deviations. Of course, these quantitative discrepancies are in part due to the fact that the elastic properties of ice I$_h$ were not taken into account in the development of any of the considered water PEFs. 

On the other hand, the qualitative discrepancies between the coarse-grained mW model and the explicit-proton PEFs provide direct insight into the molecular-level interactions in ice I$_h$. As exemplified in Figs.~\ref{Fig2} and \ref{Fig3}, a striking difference between them concerns the description of the elastic anisotropy. Without exception, all models in which the protons are explicitly taken into account substantially better capture the anisotropy of Young's and the shear moduli compared to mW, even for the TIP5P model which overestimates the stiffnesses in absolute terms by factors $\sim 2$. In their molecular description the cohesion is provided by asymmetric hydrogen bonds in which one molecule serves as a proton donor and the other as the acceptor. Furthermore, deformations can involve both translational and rotational displacements. In the mW model such rotational degrees of freedom as well as the asymmetric nature of the cohesive bond are absent, since it portrays the system as an atomic crystal. In this light, the fact that all molecular PEFs better capture the elastic anisotropy, irrespective of whether they are based on rigid or flexible molecules, suggests that it is closely related to the molecular nature of ice I$_h$, which is precisely the element that has been coarse-grained out in the mW model. The anomalous increase of the constants $C_{12}$ and $C_{13}$ with increasing temperature for the mW model is another reflection of this shortcoming, indicating that the explicit three-body term in its functional form does not adequately represent the non-central forces in ice I$_h$. 

These observations are consistent with other recent discussions concerning the limitations of the mW model in the description of mechanical properties of ice I$_h$. A recent study into the pressure-induced densification of ice I$_h$ indicates that the strong tetrahedral bias of the mW model precludes the lattice from collapsing under triaxial loading,~\cite{Guo2018} whereas the absence of explicit protons is thought to give rise to the excessive ductility of ice I$_h$ as described by the mW model.~\cite{Santos-Florez2018} These observations all highlight the molecular nature of the ice I$_h$ crystal and indicate that caution must be taken when interpreting results obtained using the  coarse-grained mW model.

\section{Summary}
\label{sec4}

In summary, using atomistic simulation techniques, we compute the elastic constants for ice I$_h$ as described by a set of 8 different PEFs for water models, including 7 explicit-proton PEFs and the coarse-grained mW potential. In the zero-temperature limit at zero pressure the the stiffness coefficients obtained using the explicit-proton models deviate substantially compared to the available experimental data. The coarse-grained mW model, on the other hand, gives results that are quantitatively closest to experiment. Calculations at finite temperature and pressure for the mW and TIP4P/Ice models display comparable behavior, with the mW model providing an overall better comparison with experiment in quantitative terms.

With respect to the qualitative behavior, however, a different picture is seen, with the discrepancies between the mW model and the explicit-proton PEFs providing insight into the molecular-level interactions in ice I$_h$. One striking difference between them involves the elastic anisotropy. While the explicit-proton models closely reproduce the qualitative trend of the experimental data, the mW model predicts a substantially more isotropic behavior. The fact that all molecular PEFs better capture the elastic anisotropy, irrespective of whether they are based on rigid or flexible molecules, suggests that it is closely related to the molecular nature of ice I$_h$, which has been coarse-grained out in the mW model. Another qualitative anomaly concerns the stiffening of the constants $C_{12}$ and $C_{13}$ with increasing temperature for the mW, which suggests that the explicit three-body term in its functional form does not adequately describe angular forces in ice I$_h$. These observations highlight the molecular nature of the ice I$_h$ crystal and indicate that caution must be taken when interpreting results obtained using the coarse-grained mW model.

\section*{Acknowledgments}

We acknowledge support from Capes, CNPq and Fapesp Grants no. 2013/08293-7, 2014/10294-4 and 2016/23891-6. The authors acknowledge the National Laboratory for Scientific Computing (LNCC/MCTI, Brazil) for providing HPC resources of the SDumont supercomputer, URL: http://sdumont.lncc.br. Part of the simulations were carried out at CCJDR-IFGW-UNICAMP, CENAPAD-SP (Brazil).

\bibliographystyle{apsrev4-1}
%

\end{document}